\documentclass[twocolumn,showpacs,aps,prl,superscriptaddress,floatfix]{revtex4}
\usepackage{graphicx}
\usepackage{multirow}
\usepackage{dcolumn}
\usepackage{bm}

\newcommand{\BABARPubYear}    {06}
\newcommand{\BABARPubNumber}  {062}
\newcommand{\SLACPubNumber} {12217}

\input pubboard/babarsym
\newcommand{\bi}{\begin{itemize}}
\newcommand{\ei}{\end{itemize}}

\newcommand{\Bztokzphig}{\ensuremath{\Bzb\to \phi\Kzb\gamma}}
\newcommand{\Bptokphig}{\ensuremath{\Bm\to \phi\Km\gamma}}
\newcommand{\Bmtokphig}{\ensuremath{\Bm\to \phi\Km\gamma}}
\newcommand{\Btokphig}{\ensuremath{\B\to \phi K\gamma}}

\newcommand{\mrec}{\ensuremath{m_{\rm rec}}}

\newcommand{\mmiss}{\ensuremath{m_{\rm miss}}}
\long\def\inst#1{\par\nobreak\kern 4pt\nobreak
    {\it #1}\par\vskip 10pt plus 3pt minus 3pt}

\begin{document}

\begin{flushleft}

\babar-PUB-\BABARPubYear/\BABARPubNumber \\
SLAC-PUB-\SLACPubNumber \\
\end{flushleft}

\title{Measurement of $B$ Decays to $\phi K \gamma$}

%% author list as of 01-Sep-2006 (603 authors)
%
\author{B.~Aubert}
\author{M.~Bona}
\author{D.~Boutigny}
\author{F.~Couderc}
\author{Y.~Karyotakis}
\author{J.~P.~Lees}
\author{V.~Poireau}
\author{V.~Tisserand}
\author{A.~Zghiche}
\affiliation{Laboratoire de Physique des Particules, IN2P3/CNRS et Universit\'e de Savoie, F-74941 Annecy-Le-Vieux, France }
\author{E.~Grauges}
\affiliation{Universitat de Barcelona, Facultat de Fisica, Departament ECM, E-08028 Barcelona, Spain }
% ...
\collaboration{The \babar\ Collaboration}
\noaffiliation

\date{\today}

\date{\today}
\begin{abstract}
We search for the decays 
$B^-\to \phi K^- \gamma$ 
and 
$\kern 0.18em\overline{\kern -0.18em B}{^0} \to \phi \kern
0.2em\overline{\kern -0.2em K}{^0} \gamma$ 
in a data sample of 228 million 
$B \kern 0.18em\overline{\kern -0.18em B}{}$ pairs collected
at the $\Upsilon(4S)$ resonance with the \mbox{\slshape
B\kern-0.1em{\smaller A}\kern-0.1em B\kern-0.1em{\smaller A\kern-0.2em
R}}\ detector.  We measure the branching fraction ${\cal B}(B^-\to
\phi K^- \gamma)~=~(3.5 \pm 0.6 \pm 0.4)\times 10^{-6}$ and set an
upper limit ${\cal B}( \kern 0.18em\overline{\kern -0.18em B}{^0} \to
\phi \kern 0.2em\overline{\kern -0.2em K}{^0} \gamma
)~<~2.7\times~10^{-6}$ at the 90\% confidence level. We also measure
the direct $C\!P$ asymmetry in $B^-\to \phi K^- \gamma$, $\calA_{C\!P}
= (-26 \pm 14 \pm 5)\%$. The uncertainties are statistical and
systematic, respectively.

\end{abstract}


\maketitle

Measurements of the branching fractions and \CP asymmetries of
\btosgam decays provide a sensitive probe of the standard model (SM), 
in which these decays are forbidden at tree level but allowed through
electroweak penguin processes.
They are sensitive to the possible effects of physics
beyond the SM manifesting as new virtual particles contributing
to loops. These additional contributions to the decay amplitudes
could affect branching fractions and \CP violation~\cite{bsgteo}. The
SM theoretical prediction \cite{Gambino:2001ew} and experimental measurements
\cite{unknown:2006bi} of the
\btosgam inclusive branching fraction have uncertainties of about 10\% and
are consistent with each other. Although exclusive \btosgam branching
fractions are experimentally easier to determine than inclusive
ones, calculations for the exclusive modes are theoretically
challenging due to large nonperturbative quantum chromodynamic effects.
The expected direct \CP asymmetry between \Bp and \Bm decay rates in the SM is
$-(0.1-1)\%$~\cite{soaresCP}, while the time-dependent
\CP asymmetry in neutral \CP eigenstates such as $\Bz \to \phi \KS
\gamma$ should be a few percent~\cite{bsgtdcp}. A significantly larger
\CP asymmetry of either type would be a sign of new physics.

There have already been results published for branching fraction and/or
\CP asymmetry measurements in several exclusive modes: $B \to \Kstar
\gamma$~\cite{kstar}, $\Bz \to \KS \piz \gamma$~\cite{kspi0g}, $B
\to \eta(') K \gamma$~\cite{ketagamma}, and various $B \to K \pi \pi
\gamma$~\cite{kpipi} modes. 
The Belle collaboration has measured
$\BR(\Bmtokphig) = (3.4 \pm 0.9 \pm 0.4) \times 10^{-6}$ and
$\BR(\Bztokzphig) < 8.3 \times 10^{-6}$ at the 90\% confidence
level using 96 million \BB pairs~\cite{Drutskoy:2003}. 
We present the first \babar\ 
measurement of the branching fraction for the charged mode \Bmtokphig\
and a search for the neutral mode \Bztokzphig~\footnote{Throughout this paper,
whenever a mode is given, the charge conjugate is also implied.} using
228 million \BB pairs.
We also measure for the first time the direct \CP asymmetry in the
charged mode $\calA_{\CP} = [N(\Bm) - N(\Bp)]/[N(\Bm) + N(\Bp)]$, 
where the flavor of the $B$ is determined by the charge of the
kaon.

The data used in this analysis were recorded with the \babar\ detector
at the \pep2\ asymmetric storage rings, in which 9.0~\gev electrons
collide with 3.1~\gev positrons to produce \FourS mesons.  The \babar\
detector is described in detail elsewhere~\cite{ref:babar}. Most important
to this analysis are the tracking system composed of the silicon
vertex tracker (SVT) and drift chamber (DCH) inside a 1.5 T magnetic
field, the ring-imaging detector of internally reflected Cherenkov
light (DIRC), and the electromagnetic calorimeter (EMC). The tracking
system can reconstruct a $B$ decay vertex with a resolution of 70~\mum
along the direction of the beam, and has a transverse momentum
resolution of 0.52\% at $500~\mevc$. The DIRC provides kaon-pion
separation of at least $4\sigma$ significance for momenta up to
3\gevc. The EMC detects photons over an energy range from 20~\mev to
9~\gev, with a resolution of 2.6\% at 2.5~\gev. A detailed Monte Carlo
(MC) simulation of signal and background processes was performed using
the {\tt EVTGEN} generator~\cite{evtgen} and the {\tt GEANT4}
package~\cite{geant4}.

We search for \Btokphig\ candidates based on charged track combinations
and the presence of a high-energy photon using a kinematic
fitter~\cite{Hulsbergen:2005pu} to reconstruct the intermediate mesons
and the $B$. Each decay vertex is required to have a \chisq probability greater
than 0.1\%. Candidates for $\phi \to \Kp\Km$ are selected from pairs of
oppositely charged tracks that have been distinguished from pions
based on a particle identification (PID) likelihood selection
algorithm that uses \dedx and Cherenkov light measurements. The same
PID algorithm is used for the single \Km from the \Bm in the charged
mode. We keep $\phi$ candidates with masses within a $\pm 10~\mevcc$
window of the nominal $\phi$ mass~\cite{Yao:2006px}. In the neutral
mode, pairs of oppositely charged tracks are accepted as \KS
candidates if they have a combined invariant mass within $\pm 10~\mevcc$
of the \KS mass and if the \KS flight length is greater than three times
its uncertainty. We require the combined $\phi K$ invariant mass
to be less than $3.0~\gevcc$. In the neutral mode a \Dz veto is applied by 
removing candidates with combined $\phi K$ invariant mass
within $\pm 10~\mevcc$ of the \Dz mass. Photon candidates are reconstructed
from EMC clusters that are not associated with charged tracks, are
isolated from other clusters, and have
the expected photon lateral shower shape. We require an energy of
$1.5-2.6~\gev$ in the \epem\ rest frame (CM frame) and we veto photon
candidates that form a \piz($\eta$) candidate with invariant mass
between $115-155~\mevcc$ ($470-620~\mevcc$) when combined with another
photon of energy greater than $50~\mev$ ($250~\mev$).

We identify signal $B$ decays through the distributions of two
quantities, missing mass and reconstructed mass, that peak around the
nominal $B$ mass. The missing mass is
$\mmiss~=~\sqrt{|p_{\FourS}~-~p_B|^2}$,  where $p_{\FourS}$ is
the \FourS four-momentum and $p_B$ is the
four-momentum of the $\Btokphig$ candidate after a mass constraint on
the $B$ is applied. The reconstructed mass \mrec\ is the $B$ candidate
invariant mass calculated from the reconstructed energy and momentum.
We require $5.12 < \mmiss < 5.32~\gevcc$ and $4.98 < \mrec < 5.48~\gevcc$.
To further discriminate $B$ decays from continuum $\epem \to \qqbar$
($q = u,d,s,c$) background we use two topological quantities: the
ratio of Legendre moments $L_2/L_0$ and the cosine of the angle
between the $B$ candidate and the \en direction in the CM frame $|\cos
\theta^*_B|$. We require $L_2/L_0 < 0.55$, where $L_i = \sum_j |p^*_j|
|\cos \theta^*_j|^i$, $p^*_j$ is the CM momentum of each particle $j$
not used in the $B$ candidate, and $\theta^*_j$ is the CM angle
between the particle's momentum and the thrust axis of the $B$
candidate. We also require $|\cos \theta^*_B| < 0.9$.

The selection criteria described above are chosen to
optimize $N_S/\sqrt{N_S + N_B}$ in the signal
region, where $N_S$ and $N_B$ are the MC simulated signal and background
yields, respectively, and the signal region is defined by {$5.05 <
\mrec < 5.4 \gevcc$, $5.27 < \mmiss < 5.29 \gevcc$,
$|\cos \theta^*_B | < 0.8$, and $L_2/L_0 < 0.48$.\label{sigregion}}
Signal MC is based on inclusive $B \to X_s \gamma$ events generated
according to the model of Kagan and Neubert~\cite{Kagan99}, using
$m_b = 4.62~\gevcc$ for the effective $b$ quark mass. Only the part of
the hadronic mass spectrum above the $\phi K$ threshold of 1.52~\gevcc
is used, with $X_s$ forced to decay to $\phi K$. This model does not
take resonances into account. 

After all criteria are applied, the average candidate multiplicity in
events with at least one candidate are 1.01 and 1.07 in
the neutral and charged modes respectively.  If
multiple $B$ candidates are found in an event, we select the best one
based on a \chisq formed from the value and uncertainty of the mass of
the $\phi$ candidate and, in the neutral mode, the \KS candidate.  The
remaining background comes from continuum combinatorics, nonresonant
$B \to K \Kp \Km \gamma$, $B \to \phi K \piz$, and $B \to \phi K
\eta$.

Signal and background yields are extracted from a fit to an unbinned
extended maximum likelihood function defined by
\begin{eqnarray}
\mathcal{L}(N_S,N_B,\vec{\alpha}) &=&  e^{-(N_S + N_B)} \times \nonumber \\ 
	& & \prod^N_i \left[N_S\mathcal{P}_S(\vec{x}_i)
	    + N_B\mathcal{P}_B(\vec{x}_i;\vec{\alpha})\right];
\label{eq:likelihood}
\end{eqnarray}
$N_S$ and $N_B$ are the number of signal and background events
respectively, the index $i$ labels each event in the data set, and $N$
is the total number of events used in the fit. $\mathcal{P}_S$ and
$\mathcal{P}_B$ are products of the one-dimensional signal and
background probability density functions (PDFs) for each of the
observables $\vec{x} = \{\mmiss,\mrec,L_2/L_0,\cos \theta^*_B\}$. The
signal shape parameters are fixed in the fit while the background parameters
$\vec{\alpha}$ are allowed to vary. In order to fit
the \CP asymmetries of signal and background in the charged mode,
the number of \Bp and \Bm events is determined separately: $N^\pm_j =
\frac{1}{2}(1 \mp \calA_{CP}^j)n_{j}$, where $j=S$ or $B$,
$n_j$ and $\calA_{CP}^j$ are the total yield and \CP asymmetry of
species $j$, respectively, and the upper (lower) signs correspond to
the positively (negatively) charged $B$ mesons.

The signal PDFs for \mmiss\ and \mrec\ are parametrized by
\begin{equation}
f(x) = \exp \left[ \frac{-x^2}{2 \sigma^2_{L,R} +
\alpha_{L,R} x^2} \right],
\end{equation}
where the parameters $\sigma_{L,R}$ and $\alpha_{L,R}$
determine the core width and variation of the width on either side of
$x=0$, $x$ being the difference from the nominal $B$ mass
of \mmiss\ or \mrec. The \mmiss\ background
PDF is an ARGUS function~\cite{argus}, with the endpoint calculated
event-by-event from the beam energy. The \mrec\ background
PDF is modeled as a $2^{\rm nd}$ degree polynomial.
The signal and background models for $L_2/L_0$ both use a binned PDF
with eight bins. The $\cos \theta^*_B$ distribution is modeled as a
$2^{\rm nd}$ degree polynomial in both signal and background; true $B$
candidates follow a $1 - \cos^2 \theta^*_B $ distribution if the
detector efficiency is flat in $\cos \theta^*_B$.

To determine the signal PDF parameters we use a  high-statistics $\Bz
\to \Kstarz(\to \Kp \pim) \gamma$ sample. Once determined, these
parameters are fixed for the fit to \Btokphig\ data. 
We determine
the selection efficiency by performing a fit of the yields on signal MC.

We apply several corrections to the signal yield and efficiency
before determining the branching fractions. Studies of simulated
events show that the main sources of signal-like (peaking) backgrounds 
are nonresonant $B \to K \Kp \Km \gamma$ events, and $B \to \phi K
\piz$ or $B \to \phi K \eta$, where one of the photons from the \piz or
$\eta$ decay is lost and the other is picked up as the signal
high-energy photon. We estimate the amount of $B \to K \Kp \Km \gamma$
contamination by fitting for the yield in $\phi$ mass sideband
regions defined by $989 < m_\phi < 1009\mevcc$ and $1029 < m_\phi <
1049\mevcc$. By interpolating into the signal region, we
find and correct for $0.0 \pm 1.5$ and $5 \pm 4$ events for the
neutral and charged modes respectively.  These contributions are
subtracted from the event yields determined in the fit. From the known
branching fraction~\cite{phikst} of $B \to \phi \Kstar(\to K \piz)$ we
correct for a contamination of $0.27 \pm 0.16$ neutral and $1.98 \pm
0.32$ charged events.
There have been no branching fraction measurements of $B \to \phi K
\piz$ or $B \to \phi K \eta$. We assume that the branching fraction of
the first is no more than one-third that of $B \to \phi \Kstar$ and
that of the latter is no more than $B \to \phi \Kstar$.
Based on this we assign an uncertainty of 0.5 neutral and
2.9 charged events due to nonresonant $B \to \phi K (\piz/\eta)$
background.  To correct for any fit bias, we generate 1000
simulated experiments using PDFs with separate
components for \BB and continuum, and embedding signal events from the
full simulation. The background components are generated using shape
parameters determined from the full MC simulation.  We correct for a
bias of $+4.1 \pm 0.5$ events in the charged mode, due to
correlations among the observables in signal MC events that are not
accounted for in the fit. In the neutral mode we find a bias of $-0.06
\pm 0.20$, so we apply no correction but include 0.20 events in the
systematic uncertainty of the yield. 
We correct for efficiency differences between
data and MC in charged track, single photon, and \KS
reconstruction. These multiplicative efficiency corrections are 0.956
in the neutral mode
and 0.975 in the charged mode. The corrected efficiencies are $(15.3
\pm 0.8)\%$ in the neutral mode and $(21.9 \pm 1.6)\%$ in the charged
mode, where the uncertainties are systematic (discussed below).

\begin{figure}
\begin{center}
\includegraphics[width=0.23\textwidth]{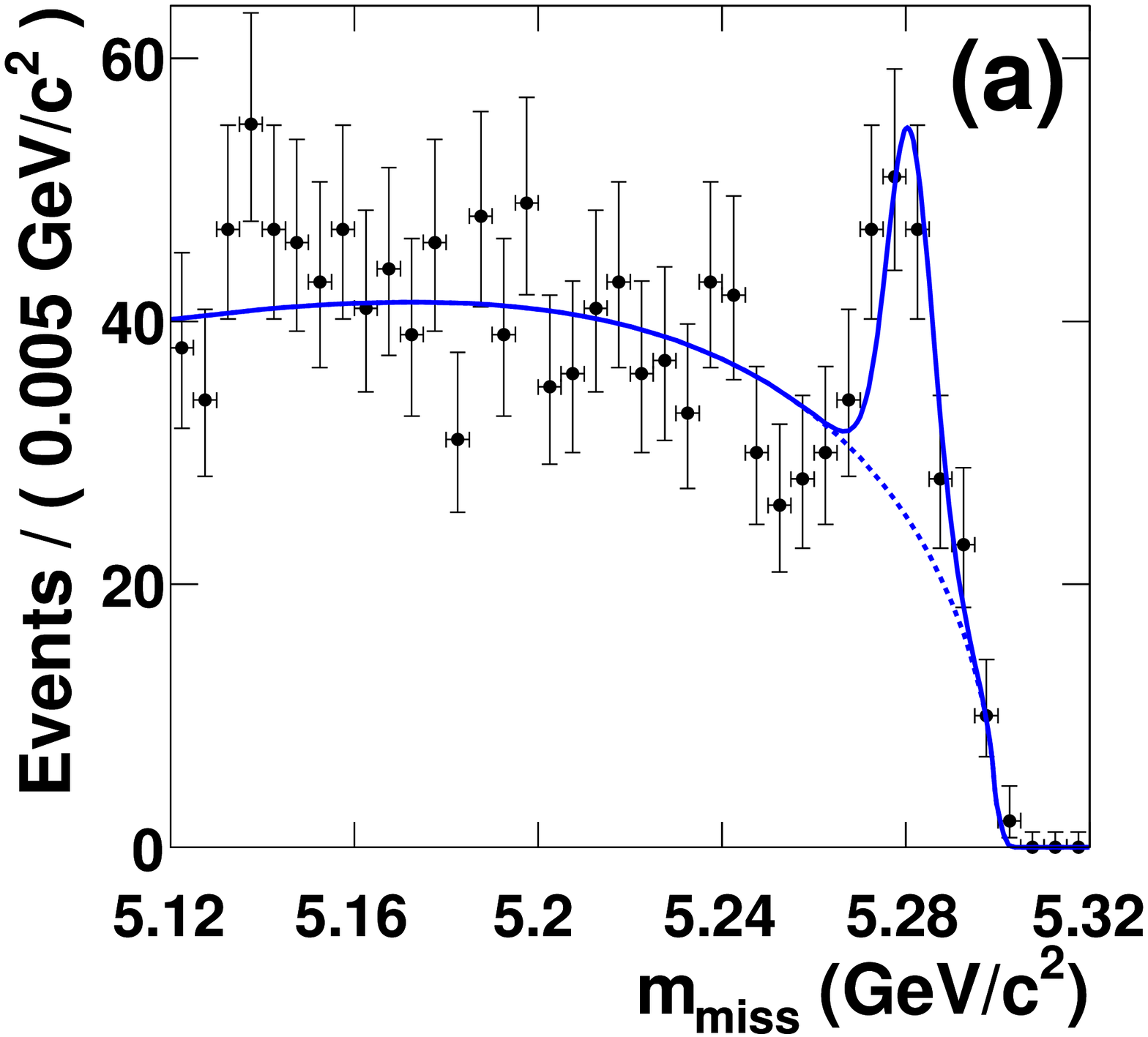}
\includegraphics[width=0.23\textwidth]{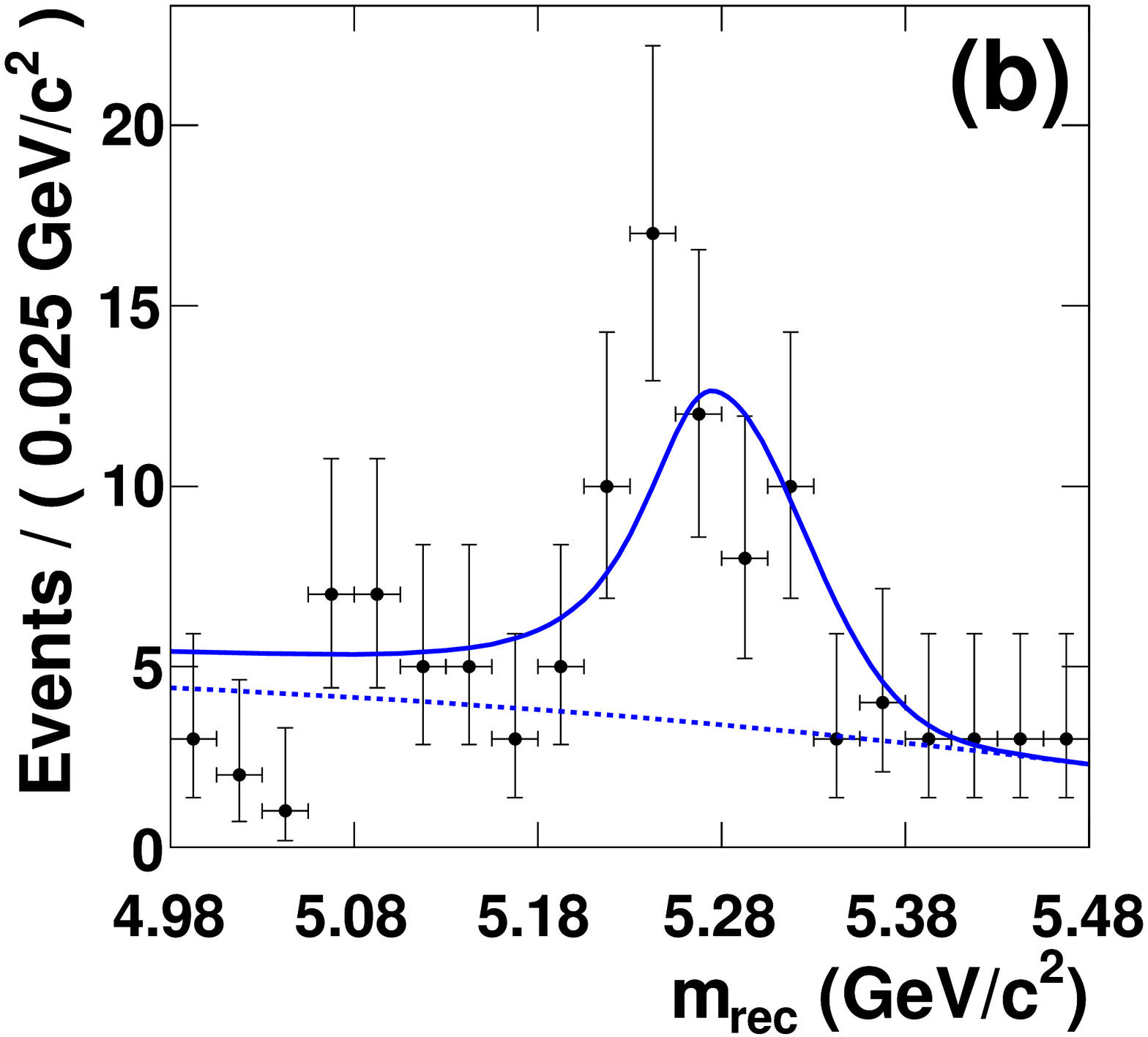}
\includegraphics[width=0.23\textwidth]{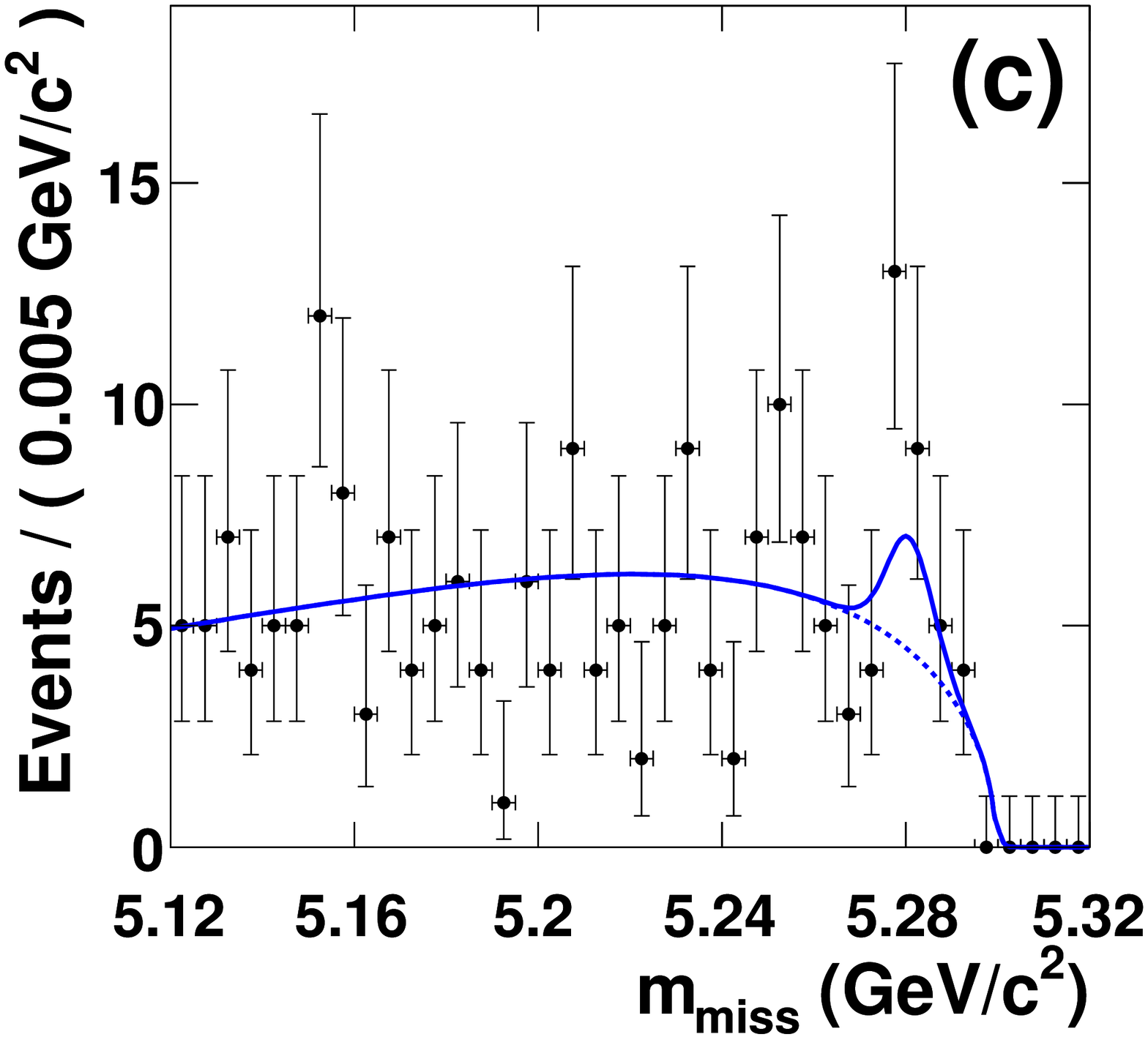}
\includegraphics[width=0.23\textwidth]{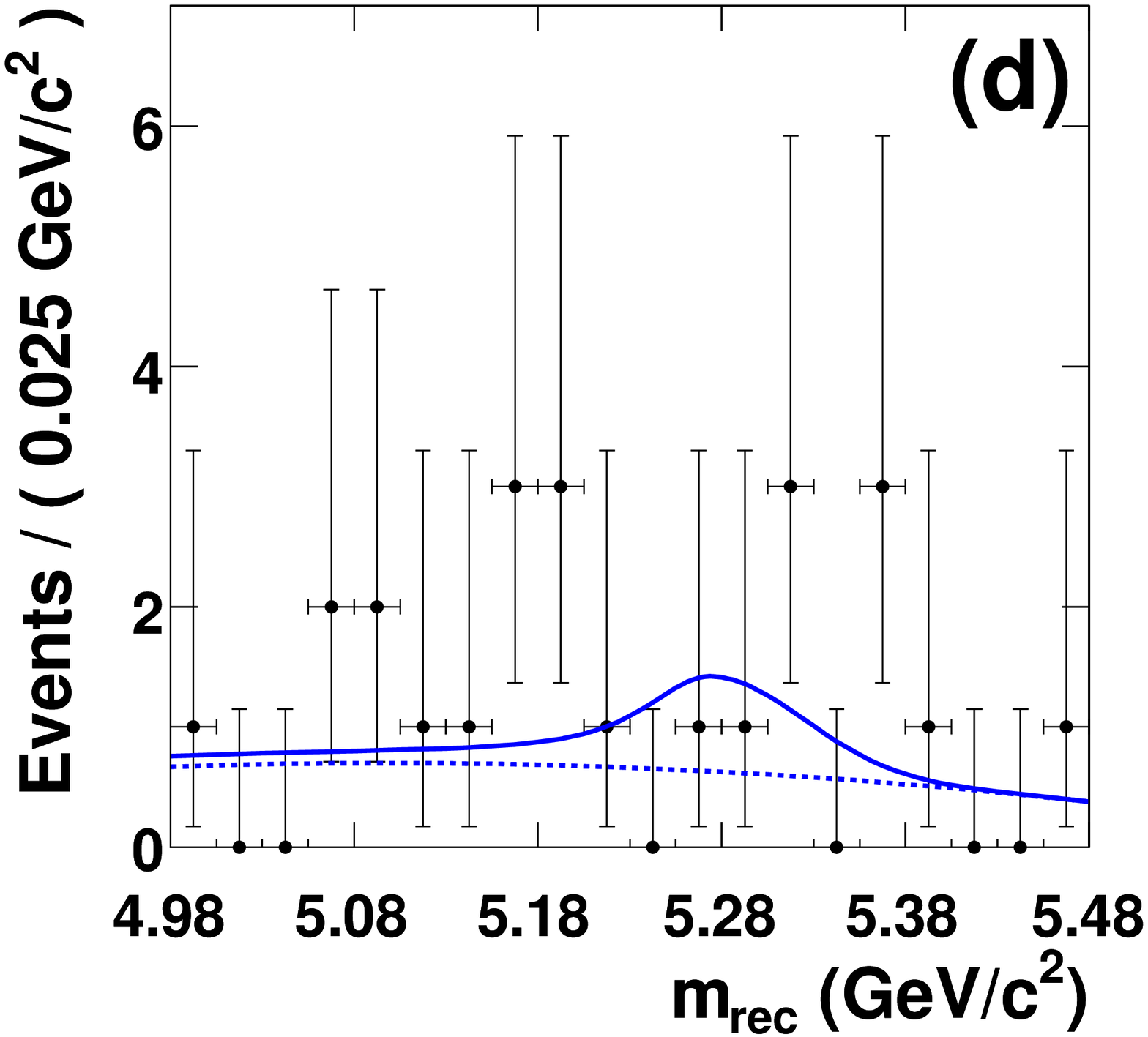}
\end{center}
\caption{Missing mass (a) and reconstructed mass (b) fits 
in the signal region for the charged mode and the
neutral mode (c,d). The dotted curves show the background
contribution while the solid curves show the sum of signal and background.
\label{fig:fitresults} }
\end{figure}

The signal yields, efficiencies, branching fractions, and charged-mode
\CP asymmetry are reported in Table~\ref{tab:final}. We calculate
the central value of the branching fractions~by
\begin{equation}
\BR_i = \frac{N^i_S}{N_{\BB} \cdot \varepsilon_i \cdot b_i}
\end{equation}
where $i$ labels either the neutral or charged mode, $N^i_S$ is the
corrected signal yield, $N_{\BB} = (228.3 \pm 2.5) \times 10^6$ is the
number of \BB pairs recorded, $\varepsilon_i$ is the
corrected efficiency, and $b_i$ is $\BR(\phi \ra \Kp \Km)[\frac{1}{2}
\BR(\KS \ra \pip \pim)]$ in the neutral mode and $\BR(\phi \to
\Kp\Km)$ in the charged mode.  The world average branching fractions
are taken from Ref.~\cite{Yao:2006px}.  We measure
$\BR(\Bmtokphig) = (3.5 \pm 0.6 \pm 0.4) \times 10^{-6}$ and
$\BR(\Bztokzphig) = (1.3 \pm 1.0 \pm 0.3) \times 10^{-6}$. In the
charged mode we measure $\calA_{\CP} = (-26 \pm 14 \pm 5)\%$. In
Fig.~\ref{fig:fitresults} we show fits to the data projected onto
\mmiss\ and \mrec. In all cases, the displayed distribution is created
with the signal region selection applied to all other fit variables.
We determine the consistency of the branching fraction measurements
with the assumption of isospin symmetry using
1000 simulated experiments in each mode with the number
of signal events determined by the average branching fraction,
$\BR_{\rm av} = 2.8 \times 10^{-6}$. From
the distribution of the differences in branching fraction between the
modes we find an 8.9\% probability to measure a difference greater
than or equal to that observed in data.

For the neutral mode we compute the 90\% confidence level upper limit
on the branching fraction. We use a Bayesian approach with a flat
prior probability for the branching fraction in the
physical region $0 \leq \BR \leq 1$ and zero elsewhere.
As the likelihood (Eq.~\ref{eq:likelihood}) is a function of
several parameters, we determine its dependence on $N_S$ by fixing $N_S$
to a series of values and recomputing the likelihood at each one,
allowing $N_B$ and $\vec{\alpha}$ to be reoptimized to obtain the
maximum likelihood at each point. We convolve this function
with a Gaussian distribution of width equal to the systematic
uncertainty of the yield. Similarly, for the efficiency
uncertainty we also use a Gaussian distribution of width equal to the
efficiency systematic uncertainty. We determine the branching
fraction upper bound $\BR_{\rm UB}$ from the following expression:
\begin{equation}
\int_{0}^{\BR_{\rm UB}}\mathcal{L(\BR)}d\BR /
\int_{0}^{1}\mathcal{L(\BR)}d\BR  = 90\%.
\end{equation}
After applying the previously discussed corrections
to the yield and efficiency, and including systematic uncertainties, we
obtain $\BR(\Bztokzphig) < 2.7 \times 10^{-6}$.
\begin{table*}
\renewcommand{\arraystretch}{1.3}
  \caption{ Summary of the branching fractions and direct \CP
asymmetry. In \Bztokzphig\ the 90\% confidence level upper limit is
also given. \label{tab:final} }
  \centerline{
    \begin{tabular}{c|cccc}
      \hline\hline
      Decay Mode      & Yield & Efficiency (\%) & $\BR (10^{-6})$ & $\calA_{CP}$ (\%)\\
      \hline
\Bptokphig  & $85 \pm 15 \pm 7$&$21.9\pm1.6\syst$ & $3.5 \pm 0.6
\pm 0.4$ & $-26 \pm 14 \pm 5$ \\
\multirow{2}{*}{\Bztokzphig} & $8 \pm 6 \pm 2$ &
\multirow{2}{*}{$15.3\pm0.8\syst$} & $1.3 \pm 1.0 \pm 0.3$ & \\
            & $<16$           &                       & $<2.7$ & \\
      \hline \hline
    \end{tabular}
  }
\end{table*}

We assign a systematic uncertainty to the yield due to the fixed
signal parameters in
the fit.  We vary these parameters within the ranges allowed by the
$\Kstar \gamma$ sample to determine the total uncertainty of the yields.
We account for other systematic uncertainties due to efficiency differences
between data and MC in charged kaon tracking, kaon PID, and \KS,
$\phi$, and photon selection. There are small uncertainties assigned
to the $L_2/L_0$ selection and the \piz / $\eta$ veto, also due to
data-MC efficiency differences.

\begin{figure}
\begin{center}
\includegraphics[width=0.30\textwidth]{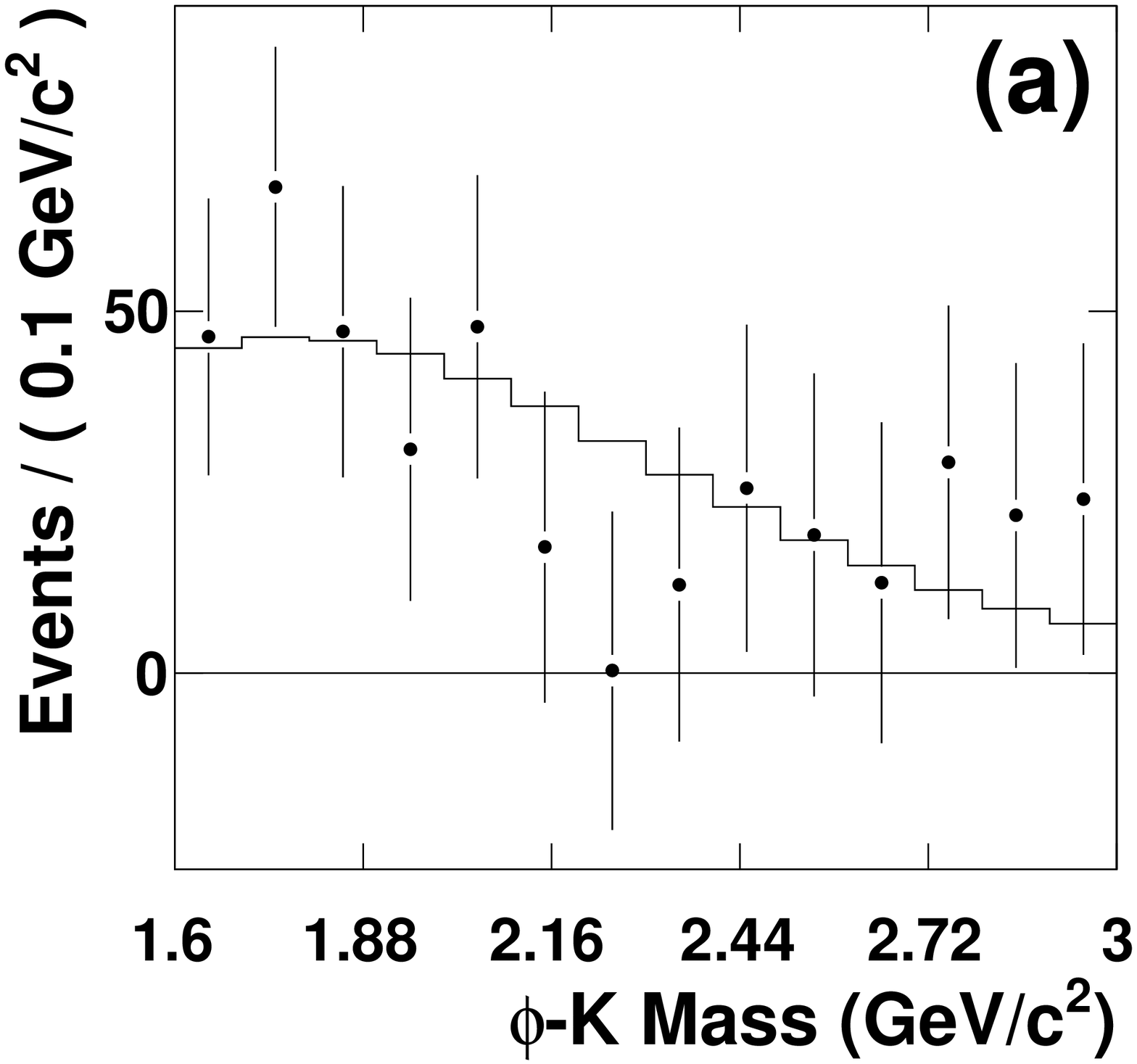}
\includegraphics[width=0.30\textwidth]{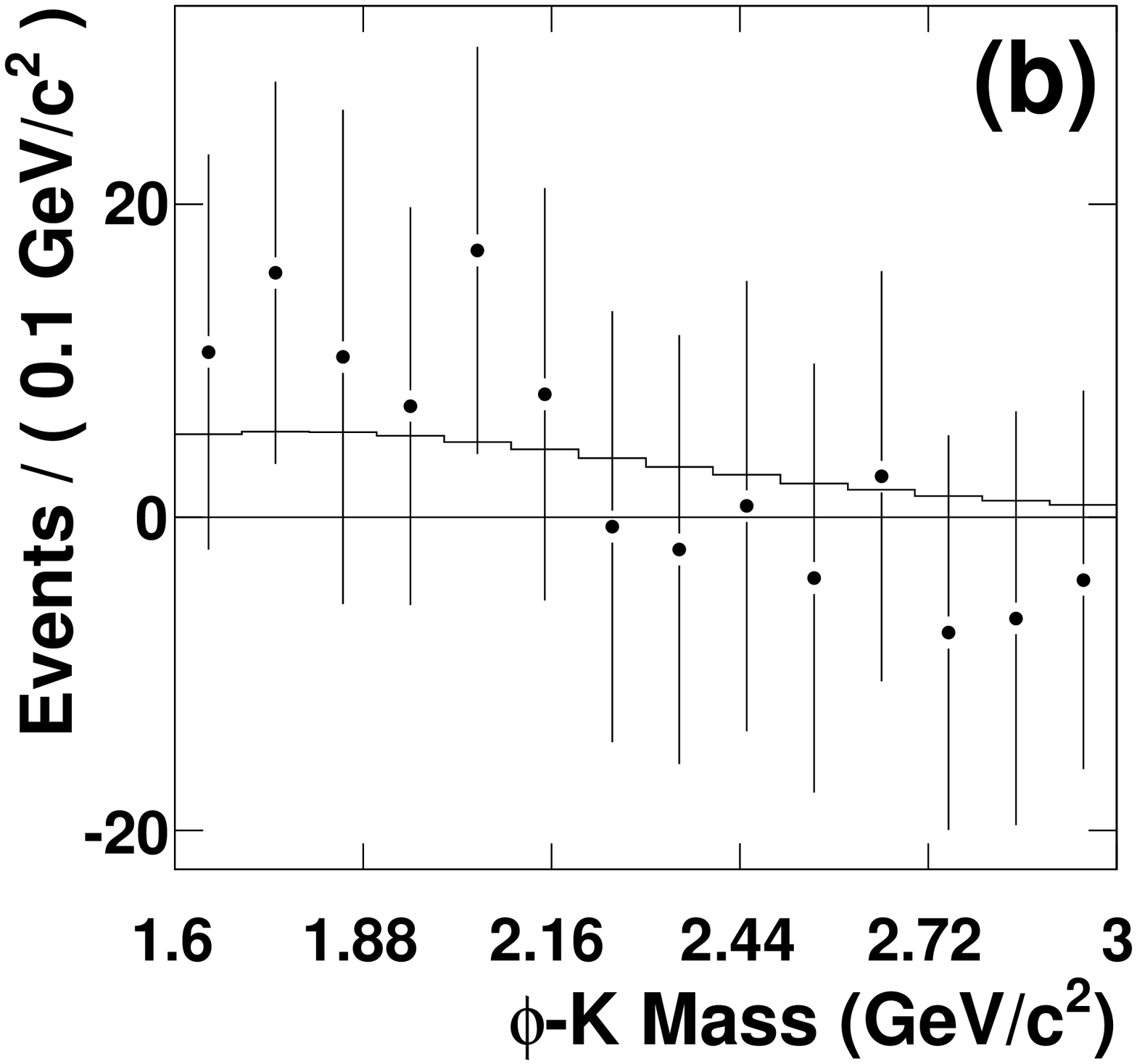}
\end{center}
\caption{The background-subtracted and efficiency-corrected $\phi K$
mass distributions
(points with uncertainties) for the charged mode (a) and the
neutral mode (b). The signal MC model for the mass spectrum, based on
Ref.~\cite{Kagan99}, is shown as a histogram without uncertainties and
is normalized to the efficiency-corrected signal yield obtained in data.
\label{fig:phikmass} }
\end{figure}

Figure~\ref{fig:phikmass} shows the efficiency-corrected $\phi K$
invariant mass distributions, using the background subtraction
technique described in Ref.~\cite{splot}. In the charged mode, we
find that no more than 50\% of the spectrum in the $1.6-3.0~\gevcc$
range can come from the $K_2(1770)$ resonance, and we use this
information to bound the uncertainty due to the assumed MC $\phi K$
mass spectrum.
We determine what the efficiency would have been
if half of the mass spectrum came from resonant $K_2(1770) \to \phi K$
production, while the other half came from the
signal MC model. We assign the relative efficiency difference between
this and the nominal model as an
uncertainty. Adding all of the previously discussed uncertainties in
quadrature, we find a total multiplicative uncertainty of 5.2\%
in the neutral mode and 7.1\% in the charged mode. The complete
systematic uncertainties for each mode are summarized in
Table~\ref{tab:syst-final}. 

\begin{table}
\renewcommand{\arraystretch}{1.3}
    \caption{Summary of the systematic uncertainties. Except where
noted, all uncertainties are given as percentages.
	\label{tab:syst-final} } 
  \centerline{
    \begin{tabular}{c|c|c}
      \hline\hline
                                   & \multicolumn{2}{c}{Uncertainty (\%)} \\
      Source                       & \Bztokzphig  & \Bmtokphig \\
      \hline
      $K\ \Kp \Km \gamma$ Subtraction  & 19.7 & 5.2 \\
      Peaking Background           & 6.4  & 3.4 \\
      Fit Bias                     & 2.6  & 0.6  \\
      Fit PDF Parameters           & $^{+7.0}_{-5.9}$  & $^{+5.9}_{-5.2}$  \\
      \hline
  \bf  Yield Uncertainty   & $\bf ^{+1.8}_{-1.7}$ events & $\bf ^{+7.3}_{-6.9}$ events \\
      \hline
      Kaon Tracking                & 2.8 & 4.2  \\
      \KS Efficiency               & 1.5 & 0  \\
      $\phi$ Efficiency            & 1.7 & 1.7  \\
      Particle ID                  & 2.8 & 4.2  \\
      Single Photon Efficiency     & 1.8 & 1.8  \\
      Photon Spectrum Model        & 0.4 & 2.6  \\
      $L_2/L_0$ Cut                & 1.2 & 1.2  \\
      $\piz / \eta$ Veto           & 1.0 & 1.0  \\
      \hline
\bf   Efficiency Uncertainty  & $\bf 5.2$ & $\bf 7.1$ \\
      \hline 
\bf   \BB Counting                 & \bf 1.1   & \bf 1.1 \\
      \hline
      \hline
\bf   Total                        & $\bf ^{+23}_{-22}$ & $\bf \pm 11$\\
      \hline \hline
    \end{tabular}}
\end{table}

For the direct \CP asymmetry measurement we bound the \Kp/\Km
efficiency asymmetry of the detector by using the measured
combinatoric background asymmetry, which is consistent with zero
within an uncertainty of 1.8\%.  To account for uncertainty due to
various peaking background sources we assume that each source can have
a \CP asymmetry of up to $\pm 58\%$, which is the root mean square
width of a flat distribution between $-1$ and 1.
We multiply this by the expected
fractional contamination in the data sample to obtain the systematic
uncertainty. For $\Bm \to \phi \Km (\piz/\eta)$ we assign 1.8\%
uncertainty, while for $\Bm \to \Km \Kp \Km \gamma$ we assign 3.5\%
uncertainty. For resonant $B \to \phi \Kstar(\to K \piz)$ events, the
previous \babar\ and Belle measurements~\cite{phikstch} show that the
\CP asymmetry is consistent with zero to within 15\%. We therefore
consider it to be negligible. As was done with the branching fraction
measurement, we vary the fixed signal parameters of the fit to obtain
a 2.2\% uncertainty for the signal \CP asymmetry. Adding the
uncertainties in quadrature we find a total $\calA_{\CP}$ systematic
uncertainty of 5\%.

In summary, we have performed the first \babar\ studies of \Btokphig\ decay
modes. We measure $\BR(\Bmtokphig) = (3.5 \pm 0.6 \pm 0.4) \times
10^{-6}$, consistent with the result from Belle. We have set
a limit $\BR(\Bztokzphig) < 2.7 \times
10^{-6}$ at the 90\% confidence level. Lastly, we have made the
first measurement of the direct \CP asymmetry in
\Bmtokphig: $\calA_{\CP} = (-26 \pm 14 \pm 5)\%$.

We are grateful for the 
extraordinary contributions of our \pep2\ colleagues in
achieving the excellent luminosity and machine conditions
that have made this work possible.
The success of this project also relies critically on the 
expertise and dedication of the computing organizations that 
support \babar.
The collaborating institutions wish to thank 
SLAC for its support and the kind hospitality extended to them. 
This work is supported by the
US Department of Energy
and National Science Foundation, the
Natural Sciences and Engineering Research Council (Canada),
Institute of High Energy Physics (China), the
Commissariat \`a l'Energie Atomique and
Institut National de Physique Nucl\'eaire et de Physique des Particules
(France), the
Bundesministerium f\"ur Bildung und Forschung and
Deutsche Forschungsgemeinschaft
(Germany), the
Istituto Nazionale di Fisica Nucleare (Italy),
the Foundation for Fundamental Research on Matter (The Netherlands),
the Research Council of Norway, the
Ministry of Science and Technology of the Russian Federation, 
Ministerio de Educaci\'on y Ciencia (Spain), and the
Particle Physics and Astronomy Research Council (United Kingdom). 
Individuals have received support from 
the Marie-Curie IEF program (European Union) and
the A. P. Sloan Foundation.

\end{document}